\documentclass[sigconf,screen]{acmart}

\usepackage{subcaption}
\usepackage{graphicx} %

\usepackage{amsmath,amsfonts}
\usepackage{array}
\usepackage{diagbox}
\usepackage{booktabs} 
\usepackage{colortbl}   
\usepackage{multirow}
\usepackage{makecell}
\usepackage{tablefootnote}
\usepackage{arydshln}
\usepackage{tikz}
\usepackage{longtable}
\usepackage{makecell}
\usepackage{pifont}
\usepackage{algorithm}
\usepackage{algpseudocode}

\algnewcommand\algorithmicinput{\textbf{Input:}}
\algnewcommand\Input{\item[\algorithmicinput]}
\algnewcommand\algorithmicoutput{\textbf{Output:}}
\algnewcommand\Output{\item[\algorithmicoutput]}

\AtBeginDocument{%
  }

\copyrightyear{2026}
\acmYear{2026}
\setcopyright{cc}
\setcctype{by}
\acmConference[SIGIR '26]{Proceedings of the 49th International ACM SIGIR Conference on Research and Development in Information Retrieval}{July 20--24, 2026}{Melbourne, VIC, Australia}
\acmBooktitle{Proceedings of the 49th International ACM SIGIR Conference on Research and Development in Information Retrieval (SIGIR '26), July 20--24, 2026, Melbourne, VIC, Australia}
\acmDOI{10.1145/3805712.3809896}
\acmISBN{979-8-4007-2599-9/2026/07}

\definecolor{azure}{rgb}{0.0, 0.5, 1.0}

\begin{document}

 
\title{Spectral Tempering for Embedding Compression in Dense Passage Retrieval}


\author{Yongkang Li}
\orcid{0000-0001-6837-6184}
\affiliation{%
  \institution{University of Amsterdam}
  \city{Amsterdam}
  \country{The Netherlands}
}
\email{y.li7@uva.nl}

\author{Panagiotis Eustratiadis}
\orcid{0000-0002-9407-1293}
\affiliation{%
  \institution{University of Amsterdam}
  \city{Amsterdam}
  \country{The Netherlands}
}
\email{p.efstratiadis@uva.nl}

\author{Evangelos Kanoulas}
\orcid{0000-0002-8312-0694}
\affiliation{%
  \institution{University of Amsterdam}
  \city{Amsterdam}
  \country{The Netherlands}
}
\email{e.kanoulas@uva.nl}


\begin{abstract}
Dimensionality reduction is critical for deploying dense retrieval systems at scale, yet mainstream post-hoc methods face a fundamental trade-off: principal component analysis (PCA) preserves dominant variance but underutilizes representational capacity, while whitening enforces isotropy at the cost of amplifying noise in the heavy-tailed eigenspectrum of retrieval embeddings. Intermediate spectral scaling methods unify these extremes by reweighting dimensions with a power coefficient $\gamma$, but treat $\gamma$ as a fixed hyperparameter that requires task-specific tuning. We show that the optimal scaling strength $\gamma$ is not a global constant: it varies systematically with target dimensionality $k$ and is governed by the signal-to-noise ratio (SNR) of the retained subspace. Based on this insight, we propose Spectral Tempering (\textbf{SpecTemp}), a learning-free method that derives an adaptive $\gamma(k)$ directly from the corpus eigenspectrum using local SNR analysis and knee-point normalization, requiring no labeled data or validation-based search. Extensive experiments demonstrate that Spectral Tempering consistently achieves near-oracle performance relative to grid-searched $\gamma^*(k)$ while remaining fully learning-free and model-agnostic. 
Our code is publicly available at \url{https://github.com/liyongkang123/SpecTemp}.
 
\end{abstract}




\maketitle
\section{Introduction}

\begin{figure}[t]
    \centering
    \includegraphics[width=\linewidth]{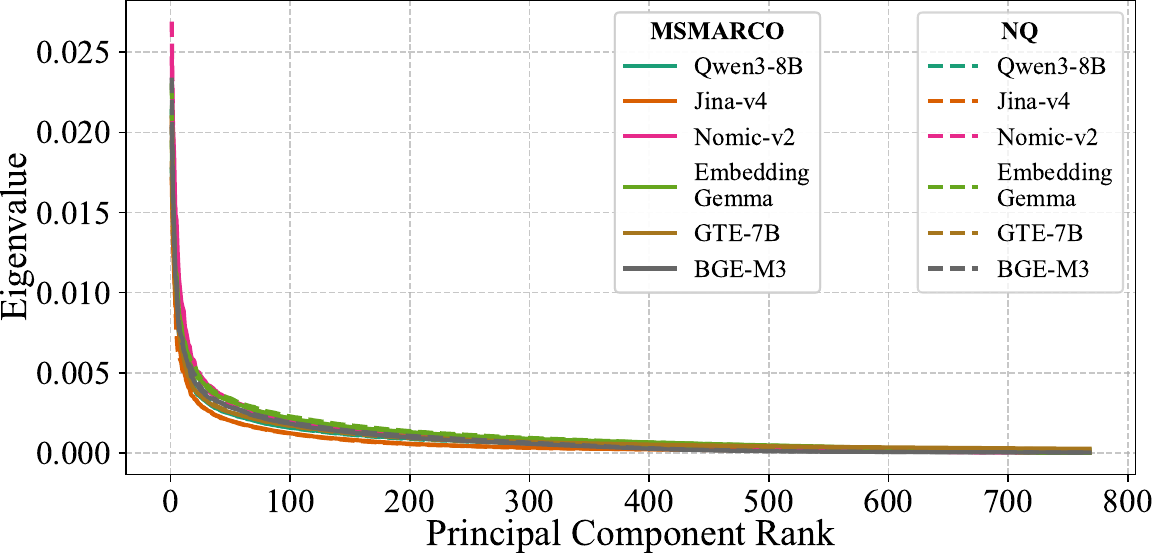}
    \caption{
    Consistent spectral structure of dense retrieval embeddings. Eigenvalue distributions from 1M sampled embeddings on MS MARCO and NQ exhibit consistent heavy-tailed decay across diverse retrievers, revealing a head–tail signal-to-noise ratio (SNR) gradient—leading components are signal-dominant while tail dimensions grow noise-prone—motivating dimensionality-adaptive tempering.\looseness=-1
    }
    \label{fig:universal_spectral_decay}
\end{figure}

Dense retrieval has become the dominant paradigm for first-stage retrieval in modern search systems~\cite{KarpukhinOMLWEC20_DPR,XiongXLTLBAO21_ANCE,reimers-2019-sentence-bert}, where queries and documents are encoded as high-dimensional embeddings and relevance is computed via similarity functions such as cosine similarity. 
While recent encoders based on Large Language Models (LLMs)~\cite{zhang2025qwen3embedding,li2023gte_7b,long2025diver} achieve state-of-the-art~(SOTA) performance, they routinely produce high-dimensional embeddings (e.g., 1024–4096), increasing the memory footprint of vector indexes and the cost of similarity computation in large-scale deployment.

To mitigate these costs, training-based approaches such as learned projections~\cite{zhang2026caseconditionawaresentence}, conditional autoencoders~\cite{liu-etal-2022-dimension}, and knowledge distillation~\cite{lioutas-etal-2020-improving} have been explored, but require retraining infrastructure tied to specific encoders.
Consequently, post-hoc compression---reducing dimensionality without parameter updates---offers a more practical alternative, yet its dominant baselines occupy flawed extremes.
Principal Component Analysis (PCA) retains maximal variance~\cite{zhang-etal-2024-evaluating-unsupervised} but leaves the energy distribution highly skewed, allowing head dimensions to overshadow complementary discriminative signals.
Conversely, standard whitening~\cite{su2021whiteningsentence} enforces isotropy by normalizing all dimensions to unit variance; yet the eigenspectrum of retrieval embeddings is heavily tailed (Figure~\ref{fig:universal_spectral_decay}), and this normalization substantially amplifies noise.
Intermediate spectral scaling methods attempt to resolve this dilemma by weighting dimensions with a fractional power $\lambda_i^{-\gamma/2}$ ($\gamma \in [0, 1]$)~\cite{kexuefm-9079}.
However, prior work treats $\gamma$ as a static hyperparameter that requires per-task tuning, overlooking that optimal tempering varies systematically with the target dimensionality $k$.
For instance, aggressive whitening ($\gamma \approx 1$) benefits compact subspaces ($k=64$) but degrades quality at large $k$ by amplifying low SNR tail components.

In this work, we formalize this dimensionality-dependent behavior through a local SNR analysis of the corpus eigenspectrum. By estimating a spectral noise floor, we obtain an SNR profile that reveals a smooth head–tail transition from signal-dominant to noise-prone components—explaining why optimal tempering strength should decrease as target dimensionality $k$ grows to include low-SNR tail directions. Building on this insight, we propose
Spectral Tempering (\textbf{SpecTemp}), a learning-free method that analytically derives an adaptive $\gamma(k)$ directly from the SNR profile, automatically interpolating between variance preservation (PCA) and isotropy (whitening). The resulting linear transform is computed offline from corpus embeddings and applied identically to queries at inference time, requiring no labeled data or validation tuning.

Our contributions are three-fold:

$\bullet$ We characterize the  \emph{dimensionality-dependent} optimality of spectral scaling, demonstrating that the ideal $\gamma$ is intrinsically governed by the subspace SNR rather than being a fixed constant.

$\bullet$ We propose \textbf{SpecTemp}, a learning-free method that analytically derives an adaptive $\gamma(k)$ from the corpus eigenspectrum, requiring no labeled data or validation-based tuning.

$\bullet$ We conduct extensive experiments across multiple LLM-based embedding models and diverse retrieval datasets, demonstrating that SpecTemp consistently achieves near-oracle performance relative to grid-searched $\gamma^*(k)$.

\section{Related Work}

\paragraph{Dense Retrieval.}
Dense retrieval has evolved from BERT-based bi-encoders~\cite{DevlinCLT19_bert,KarpukhinOMLWEC20_DPR,XiongXLTLBAO21_ANCE,HofstatterLYLH21_tasb_dense_retrieval,LinALOLMY023_dragon} with compact 768d representations to massive LLM-based architectures.
To capture complex semantics, recent SOTA models like RepLLaMA~\cite{MaWYWL24_RankLLaMA}, E5-Mistral~\cite{WangE5-base}, and Qwen3-Embedding~\cite{zhang2025qwen3embedding} employ billion-scale, often decoder-only backbones.
While yielding superior generalization, this shift often produces high-dimensional embeddings (e.g., 4096d), creating the storage bottlenecks that motivate our study. 
Recent work has also examined dense retrievers beyond effectiveness and efficiency, including their robustness under adversarial settings~\cite{li2026understanding}, such as query perturbations~\cite{PenhaCH22_query_variation} and corpus poisoning~\cite{ZhongHWC23_Poisoning,10.1145/3583780.3614793_MCARA,li2025reproducinghotflip,li2025unsupervised}.

\paragraph{Embedding Compression.}
Strategies to mitigate these overheads fall into two broad categories: training-based and post-hoc.

\textbf{Training-based methods} optimize compression objectives during or after training-time. 
Matryoshka Representation Learning (MRL)~\cite{KusupatiBRWSRHC22_MRL} has gained widespread adoption for enabling flexible truncation by nesting information in prefix dimensions. 
Other approaches employ knowledge distillation to transfer capabilities to smaller students~\cite{lioutas-etal-2020-improving}, or optimize conditional autoencoders to compress fixed embeddings into latent codes~\cite{liu-etal-2022-dimension}. 
While effective, these strategies require additional training data and incur high computational costs for retraining, rendering them impractical for off-the-shelf or API-only models.

\textbf{Post-hoc methods}, in contrast, transform pretrained embeddings without parameter updates. 
Spectral projections dominate this landscape, scaling dimensions based on their eigenvalues.
PCA ($\gamma=0$) maximizes variance but leaves the space anisotropic~\cite{zhang-etal-2024-evaluating-unsupervised,ma-etal-2021-simple, zuo2026efficiencyembeddingcompress}, while Standard Whitening ($\gamma=1$) enforces isotropy but risks amplifying tail noise~\cite{su2021whiteningsentence,huang-etal-2021-whiteningbert-easy}.
Intermediate strategies employ a fractional exponent $\gamma \in [0,1]$ to interpolate between these extremes~\cite{kexuefm-9079}, yet they rely on a static hyperparameter requiring per-task tuning.
Alternatively, Random Projection offers dimension-agnostic compression via the Johnson--Lindenstrauss lemma~\cite{johnson1984extensions} but ignores the learned manifold structure.

A separate line of work targets isotropy via post-processing, such as removing dominant directions~\cite{mu2018allbutthetop,rajaee-pilehvar-2021-cluster,raunak-etal-2019-effective} or mapping to uniform distributions~\cite{li-etal-2020-sentence}, though these focus on quality rather than dimensionality reduction.
Similarly, Product Quantization (PQ)~\cite{jegou2010product} and its variants achieve index-level compression via codebooks~\cite{douze2024faiss}; being a downstream operation, this approach is orthogonal to and composable with linear projections like ours.

\textbf{SpecTemp} occupies a distinct position in this landscape: it is a \textit{post-hoc, learning-free} linear projection that derives a dimensionality-adaptive tempering strength $\gamma(k)$ from the local SNR of the retained subspace, requiring no labeled data, retraining, validation-based tuning, or index-level modifications.

\section{Methodology}

We now describe Spectral Tempering~(\textbf{SpecTemp}), a post-hoc compression method that derives a dimensionality-adaptive tempering exponent $\gamma(k)$ directly from the eigenspectrum of corpus embeddings. The method proceeds in three stages: spectral decomposition, SNR-guided exponent derivation, and embedding transformation.

\subsection{Spectral Decomposition}

Given a corpus embedding matrix $\mathbf{X} \in \mathbb{R}^{n \times d}$, we first center it by subtracting the column-wise mean $\boldsymbol{\mu}$:
\begin{equation}
    \bar{\mathbf{X}} = \mathbf{X} - \mathbf{1}\boldsymbol{\mu}^\top
\end{equation}
Centering reduces the influence of a global offset direction and yields a more stable covariance spectrum; we apply the same corpus-derived centering to both queries and documents to preserve geometric consistency.
We then compute the eigendecomposition of the covariance matrix:
\begin{equation}
    \mathbf{C} = \frac{1}{n-1}\bar{\mathbf{X}}^\top \bar{\mathbf{X}} = \mathbf{U} \boldsymbol{\Lambda} \mathbf{U}^\top
\end{equation}
where $\boldsymbol{\Lambda} = \mathrm{diag}(\lambda_1, \dots, \lambda_d)$ with $\lambda_1 \ge \lambda_2 \ge \dots \ge \lambda_d$, and $\mathbf{U} = [\mathbf{u}_1, \dots, \mathbf{u}_d]$ are the corresponding eigenvectors.

\subsection{SNR-Guided Exponent Derivation}

The core insight of Spectral Tempering is that the appropriate tempering strength should be governed by the signal quality of the retained subspace. We formalize this through a local SNR analysis.

\paragraph{Noise Floor Estimation.}
We estimate the noise floor $\sigma^2_{\text{noise}}$ as the mean eigenvalue of the spectral tail:
\begin{equation}\label{eq:noise_floor}
    \sigma^2_{\text{noise}} = \frac{1}{|\mathcal{T}|} \sum_{i \in \mathcal{T}} \lambda_i
\end{equation}
where $\mathcal{T}$ denotes the last 10\% of eigenvalue indices. 
As shown in Figure~\ref{fig:universal_spectral_decay}, diverse retrieval encoders exhibit a consistently heavy-tailed eigenspectrum whose tail consistently plateaus into a stable noise floor, making this region a reliable, model-agnostic anchor for noise estimation.
We verify in Section~\ref{sec:sensitivity_analysis} that SpecTemp is insensitive to the exact percentile choice, confirming that this default requires no per-task tuning.

\paragraph{Local SNR Computation.}
The local SNR at rank $i$ measures the excess energy above the noise floor:
\begin{equation}
    \mathrm{SNR}(i) = \max\!\left(0,\; \frac{\lambda_i - \sigma^2_{\text{noise}}}{\sigma^2_{\text{noise}}}\right)
\end{equation}
We note that this quantity is not intended as a generative statistical estimate in the sense of spiked covariance models, but as a monotonic, spectrum-level proxy for relative signal dominance---sufficient for calibrating the tempering exponent. This quantity is large for head components where the signal dominates, and vanishes in the tail where eigenvalues converge to the noise floor.

\paragraph{Anchor Point and Adaptive $\gamma(k)$.}
To derive $\gamma(k)$ without task-specific tuning, we need a reference point that separates the high-confidence signal regime from the transitional regime. We identify this anchor as the \textit{knee point} of the SNR curve---the rank at which SNR transitions from rapid to gradual decay---detected via the Kneedle algorithm\footnote{\url{https://github.com/arvkevi/kneed}}~\cite{SatopaaAIR11_Kneedle}. Let $k_{\text{knee}}$ denote this rank and $S_{\text{ref}} = \mathrm{SNR}(k_{\text{knee}})$ the corresponding SNR value.

Since the $k$-th component defines the noise bottleneck of the retained subspace, we use its SNR as a conservative proxy for subspace signal quality. This ensures that the tempering strength is constrained by the worst-case noise exposure rather than being overly influenced by optimistic, high-variance directions. The adaptive exponent for target dimensionality $k$ is then:
\begin{equation}\label{eq:gamma}
    \gamma(k) = \min\!\left(1,\; \frac{\mathrm{SNR}(k)}{S_{\text{ref}}}\right)
\end{equation}
Normalizing by $S_{\text{ref}}$ ensures that all target dimensionalities within the high-SNR regime ($k \le k_{\text{knee}}$) receive full whitening ($\gamma = 1$), while dimensions beyond the knee are progressively tempered. 
We adopt a linear mapping between SNR and $\gamma$ following the principle of parsimony, as this simple formulation avoids introducing additional degrees of freedom and is empirically sufficient and robust.
The resulting behavior is intuitive: for small target dimensions (e.g., $k{=}64$), the retained subspace lies entirely in the high-SNR regime, yielding $\gamma(k)\approx 1$ (near-whitening). As $k$ increases toward the full dimensionality, the subspace increasingly includes noise-dominated components, and $\gamma(k)$ gracefully decreases toward 0 (near-PCA).

\subsection{Transformation}

Given target dimensionality $k$, we construct the transformation matrix by combining the top-$k$ eigenvectors with the derived exponent:\looseness=-1
\begin{equation}
    \mathbf{W}_k = \mathbf{U}_k \cdot \mathrm{diag}\!\left(\lambda_1^{-\gamma(k)/2},\; \dots,\; \lambda_k^{-\gamma(k)/2}\right)
\end{equation}
where $\mathbf{U}_k = [\mathbf{u}_1, \dots, \mathbf{u}_k] \in \mathbb{R}^{d \times k}$. The compressed embedding for any input $\mathbf{x}$ (query or document) is:
\begin{equation}
    \mathbf{y} = (\mathbf{x} - \boldsymbol{\mu})^\top \mathbf{W}_k \in \mathbb{R}^k
\end{equation}
The eigendecomposition is computed once on a corpus sample; the resulting $\boldsymbol{\mu}$ and $\mathbf{W}_k$ are then applied identically to documents (offline) and queries (online), ensuring compatibility with standard ANN indexing.  When the downstream similarity metric is cosine similarity, the transformed vectors are additionally L2-normalized.

\section{Experiments}

In this section, we present empirical evaluations to validate the effectiveness of \textbf{SpecTemp} across diverse retrieval datasets.

\subsection{Experiment Setup}
\subsubsection{Datasets}
We evaluate on four retrieval datasets: \textbf{MS MARCO} Passage Ranking~\cite{bajaj2016ms} for web search, Natural Questions (\textbf{NQ})~\cite{KwiatkowskiPRCP19_NQ} for open-domain QA, \textbf{FEVER}~\cite{ThorneVCM18_FEVER} for evidence retrieval in fact verification, and \textbf{FiQA}~\cite{MaiaHFDMZB18_fiqa} for domain-specific financial retrieval, 
covering diverse domains and scales.   
These datasets span diverse query types, corpus scales, and domains.

\subsubsection{Retrieval Models}
We experiment on six widely used open-source dense retrievers spanning different scales and embedding dimensions, as summarized in Table~\ref{tab:models}.
Four models (\textbf{Qwen3-8B}~\cite{zhang2025qwen3embedding}\footnote{\url{https://huggingface.co/Qwen/Qwen3-Embedding-8B}}, \textbf{Jina-v4}~\cite{günther2025jinaembeddingsv4}\footnote{\url{https://huggingface.co/jinaai/jina-embeddings-v4}}, \textbf{Nomic-v2}~\cite{nussbaum2025_nomicv2}\footnote{\url{https://huggingface.co/nomic-ai/nomic-embed-text-v2-moe}}, \textbf{EmbeddingGemma}~\cite{vera2025embeddinggemma}\footnote{\url{https://huggingface.co/google/embeddinggemma-300m}}) support Matryoshka Representation Learning, providing strong truncation baselines. Two models (\textbf{GTE-7B}~\cite{li2023gte_7b}\footnote{\url{https://huggingface.co/Alibaba-NLP/gte-Qwen2-7B-instruct}}, \textbf{BGE-M3}~\cite{chen-etal-2024-m3}\footnote{\url{https://huggingface.co/BAAI/bge-m3}}) lack native MRL support, testing the generality of post-hoc compression.

\begin{table}[t]
\centering
\caption{Retrieval model statistics (Params: total parameters; MRL: Matryoshka Representation Learning support).}
\label{tab:models}
\setlength{\tabcolsep}{3.4pt}
\renewcommand\arraystretch{1.1}
\begin{tabular}{lccccc}
\hline
\textbf{Model} & \textbf{Params} & \textbf{Dim} &\textbf{Length} & \textbf{MRL} & \textbf{Release} \\ \hline
Qwen3-8B       & 8.0B            & 4096    &32k      & \checkmark   & Jun 2025         \\
Jina-v4        & 3.8B            & 2048 &32k        & \checkmark   & Jun 2025         \\
Nomic-v2       & 475M            & 768  & 512        & \checkmark   & Feb 2025         \\
EmbeddingGemma      & 308M            & 768  &2048         & \checkmark   & Sep 2025         \\ 
\hline
GTE-7B         & 7.0B            & 3584   &32k      & \ding{55}    & Jun 2024         \\
BGE-M3          & 560M           & 1024   &8192      & \ding{55}    & Feb 2024         \\ \hline
\end{tabular}
\begin{flushleft}
\end{flushleft}
\vspace{-0.3cm}
\end{table}

\subsubsection{Baselines}
We compare against representative learning-free post-hoc methods that require no labeled data or model fine-tuning. \textbf{Prefix Truncation} retains the first $k$ dimensions (standard for MRL-compatible models). \textbf{Random Truncation} subsamples $k$ dimensions, a simple strategy shown to be surprisingly competitive in recent work~\cite{takeshita_2025_randomly50}. \textbf{Random Projection} compresses via a Gaussian random matrix as a theoretical baseline. For spectral methods, we evaluate \textbf{PCA} ($\gamma=0$), \textbf{Standard Whitening} ($\gamma=1$), and \textbf{$\gamma$-Whitening} with a fixed $\gamma=0.5$ to represent static power normalization. 
All spectral baselines, including ours, fit their transformations on the document corpus. We exclude quantization-based methods (e.g., PQ) since they are complementary to our learning-free spectral post-processing and can be combined with SpecTemp; we focus on isolating the effect of embedding-space compression. 


\begin{table*}[t]
\centering
 
\caption{Retrieval performance on four datasets at target dimensions $k$.
Bold denotes the best per column within each model.
All results are averaged over three random seeds (1999, 5, 2026).
Superscript $^\text{ns}$ indicates that, for all three runs, the difference from \textit{Full Dimension} is not significant (two-sided paired $t$-test, $p<0.05$). Absence of $^\text{ns}$ indicates significance in at least one run.}

\label{tab:main_results}
\small
\setlength{\tabcolsep}{1.5pt}
\renewcommand\arraystretch{1}
\begin{tabular}{l|llllll:lllll:lllll:llllllll} 
\hline
\multirow{2}{*}{\textbf{Model}} & \multirow[b]{2}{*}{\raisebox{-0.8ex}{\makecell[c]{\makebox[2.2cm][c]{\textbf{Method} $\downarrow$ \hfill $k$ $\rightarrow$}}}}

 & \multicolumn{5}{c}{\textbf{MS MARCO} } & \multicolumn{5}{c}{\textbf{NQ} } & \multicolumn{5}{c}{\textbf{FEVER} } & \multicolumn{5}{c}{\textbf{FiQA} }\\
\cmidrule(lr){3-7} \cmidrule(lr){8-12} \cmidrule(lr){13-17} \cmidrule(lr){18-22}
 & & \textbf{768} & \textbf{512} & \textbf{256} & \textbf{128} & \textbf{64} & \textbf{768} & \textbf{512} & \textbf{256} & \textbf{128} & \textbf{64} & \textbf{768} & \textbf{512} & \textbf{256} & \textbf{128} & \textbf{64} & \textbf{768} & \textbf{512} & \textbf{256} & \textbf{128} & \textbf{64} \\ \hline

\multirow{8}{*}{Qwen3-8B} 
& {Full Dimension} & \multicolumn{5}{c}{\cellcolor{gray!30}\textbf{36.8}} & \multicolumn{5}{c}{\cellcolor{gray!30}\textbf{64.9}} & \multicolumn{5}{c}{\cellcolor{gray!30}\textbf{91.8}} & \multicolumn{5}{c}{\cellcolor{gray!30}\textbf{64.7}}\\
 & Prefix Truncation & \textbf{36.4} & 35.7 & 34.5 & 32.4 & \textbf{28.2} & 63.7 & 63.2 & 61.1 & 57.3 & 49.1 & \textbf{91.8}$^{\text{ns}}$ & \textbf{91.6} & \textbf{91.2} & \textbf{90.1} & \textbf{85.5} & 63.9 & 63.5 & 61.6 & 57.4 & 51.3 \\
 & Random Truncation & 35.9 & 35.5 & 34.3 & 31.5 & 24.8 & 63.5 & 62.5 & 59.3 & 53.2 & 40.0 & 91.6$^{\text{ns}}$ & 91.3 & 90.8 & 89.0 & 79.6 & 63.0 & 61.6 & 59.4 & 53.4 & 41.7 \\
 & Random Projection & 36.2 & \textbf{35.7} & 34.3 & 32.0 & 26.4 & 63.6 & 62.5 & 60.4 & 55.5 & 44.2 & 91.5 & 91.3 & 90.6 & 89.4 & 82.3 & 63.0 & 62.6 & 59.8 & 55.5 & 45.5 \\
 & PCA & 36.0 & 35.5 & 34.1 & 31.3 & 25.0 & 64.6$^{\text{ns}}$ & 63.8 & 62.1 & 57.2 & 47.0 & 91.0 & 90.7 & 89.6 & 87.4 & 83.1 & 63.7 & 63.7 & 62.1 & 59.2 & 53.2 \\
 & Whitening & 34.9 & 35.2 & 34.1 & 32.3 & 26.9 & 61.9 & 62.3 & 61.8 & 58.3 & \textbf{49.4} & 91.0 & 90.8 & 89.8 & 87.1 & 83.5 & 58.4 & 59.8 & 60.5 & 59.0 & 52.8 \\
 & $\gamma$-Whitening & 35.9 & 35.5 & 34.6 & 32.3 & 26.4 & 64.1 & 63.9 & 62.5 & \textbf{58.8} & 49.1 & 91.1 & 91.0 & 89.9 & 87.6 & 83.6 & 62.6 & 62.4 & 62.5 & \textbf{59.8} & \textbf{53.8} \\
 & \textbf{SpecTemp} & 36.1 & 35.6 & \textbf{34.6} & \textbf{32.4} & 26.8 & \textbf{64.9}$^{\text{ns}}$ & \textbf{64.1} & \textbf{62.6} & 58.7 & \textbf{49.4} & 91.2 & 90.9 & 89.9 & 87.2 & 83.5 & \textbf{64.0} & \textbf{63.7} & \textbf{62.8} & 59.7 & 53.1 \\
 \hline

\multirow{8}{*}{Jina-v4} 
& {Full Dimension} & \multicolumn{5}{c}{\cellcolor{gray!30}\textbf{32.1}} & \multicolumn{5}{c}{\cellcolor{gray!30}\textbf{61.6}} & \multicolumn{5}{c}{\cellcolor{gray!30}\textbf{87.8}} & \multicolumn{5}{c}{\cellcolor{gray!30}\textbf{47.7}}\\
 & Prefix Truncation & 31.4 & 31.1 & 30.2 & 28.1 & 21.9 & 60.8 & 60.3 & 57.8 & 53.3 & 41.5 & 87.4 & \textbf{87.3} & 85.8 & 82.2 & 67.0 & 47.0$^{\text{ns}}$ & 46.8$^{\text{ns}}$ & 44.3 & 40.6 & 31.1 \\
 & Random Truncation & 31.4 & 30.9 & 29.4 & 26.7 & 20.1 & 60.4 & 59.4 & 56.1 & 50.1 & 35.9 & 87.0 & 86.4 & 84.1 & 78.4 & 60.9 & 46.4 & 45.2 & 42.6 & 37.2 & 27.7 \\
 & Random Projection & 31.1 & 30.7 & 29.2 & 26.5 & 20.8 & 59.9 & 59.4 & 56.3 & 50.7 & 38.1 & 86.8 & 86.1 & 84.7 & 78.4 & 63.4 & 46.0 & 45.5 & 43.4 & 38.0 & 28.3 \\
 & PCA & 31.9$^{\text{ns}}$ & 31.5 & 30.4 & 27.5 & 18.6 & 61.9$^{\text{ns}}$ & 61.6$^{\text{ns}}$ & 60.1 & 55.8 & 43.0 & 87.4 & 87.0 & 85.4 & 81.5 & 69.0 & 46.8 & 46.7 & 45.6 & 43.1 & 37.7 \\
 & Whitening & 29.0 & 30.0 & 30.7 & \textbf{29.8} & \textbf{24.2} & 56.2 & 57.1 & 57.8 & 56.4 & 49.6 & 84.9 & 85.1 & 84.7 & 82.0 & 71.4 & 41.0 & 41.6 & 42.4 & 42.1 & 36.8 \\
 & $\gamma$-Whitening & 31.3 & 31.7$^{\text{ns}}$ & \textbf{31.4} & 29.5 & 22.7 & 60.3 & 60.6 & 60.5 & 58.2 & 49.3 & 87.1 & 87.0 & \textbf{86.0} & \textbf{82.6} & 71.7 & 45.7 & 45.7 & 45.2 & 43.9 & \textbf{37.8} \\
 & \textbf{SpecTemp} & \textbf{31.9}$^{\text{ns}}$ & \textbf{31.8} & 31.2 & 29.3 & 23.7 & \textbf{62.1} & \textbf{61.7}$^{\text{ns}}$ & \textbf{61.0} & \textbf{58.3} & \textbf{49.8} & \textbf{87.5} & 87.1 & 85.8 & 82.6 & \textbf{71.7} & \textbf{47.2}$^{\text{ns}}$ & \textbf{47.0} & \textbf{45.6} & \textbf{43.9} & 37.2 \\
 \hline

\multirow{8}{*}{GTE-7B} 
& {Full Dimension} & \multicolumn{5}{c}{\cellcolor{gray!30}\textbf{39.1}} & \multicolumn{5}{c}{\cellcolor{gray!30}\textbf{66.8}} & \multicolumn{5}{c}{\cellcolor{gray!30}\textbf{95.2}} & \multicolumn{5}{c}{\cellcolor{gray!30}\textbf{61.8}}\\
 & Prefix Truncation & 38.4 & 38.0 & 36.9 & 34.2 & 28.7 & 65.3 & 64.7 & 62.0 & 57.0 & 47.2 & 95.0 & 95.0 & 94.5 & 93.8 & 91.2 & 59.8 & 58.5 & 53.6 & 48.6 & 40.3 \\
 & Random Truncation & 38.4 & 37.9 & 36.5 & 34.1 & 28.3 & 65.3 & 64.4 & 61.4 & 56.9 & 45.1 & 94.9$^{\text{ns}}$ & 94.6 & 94.3 & 93.4 & 89.3 & 60.3 & 59.4 & 56.6 & 50.3 & 39.4 \\
 & Random Projection & 38.3 & 37.8 & 36.7 & 34.4 & 29.1 & 65.3 & 64.4 & 62.5 & 57.3 & 47.6 & 94.8 & 94.8 & 94.4 & 93.5 & 91.1 & 60.6 & 59.7 & 56.7 & 52.1 & 41.7 \\
 & PCA & 38.8 & 38.3 & 36.9 & 34.7 & 29.9 & 67.1$^{\text{ns}}$ & 66.4 & 64.3 & 59.7 & 50.4 & 95.2$^{\text{ns}}$ & 95.1$^{\text{ns}}$ & 94.7 & 93.8 & 90.6 & 62.3$^{\text{ns}}$ & 61.6$^{\text{ns}}$ & 58.8 & 55.9 & 48.7 \\
 & Whitening & 37.2 & 37.4 & 36.6 & 35.0 & \textbf{31.0} & 64.4 & 65.2 & 64.5 & 61.1 & \textbf{52.9} & \textbf{95.5} & 95.3 & 95.0$^{\text{ns}}$ & \textbf{94.3} & \textbf{92.4} & 58.5 & 59.6 & 59.3 & 57.0 & \textbf{51.3} \\
 & $\gamma$-Whitening & 38.2 & 38.3 & 37.0 & \textbf{35.2} & 30.7 & 66.4 & 66.6$^{\text{ns}}$ & \textbf{65.1} & 61.3 & 52.4 & 95.5 & \textbf{95.4} & \textbf{95.0}$^{\text{ns}}$ & 94.3 & 92.0 & 62.1$^{\text{ns}}$ & 62.0$^{\text{ns}}$ & \textbf{60.5} & 57.1 & 50.7 \\
 & \textbf{SpecTemp} & \textbf{38.9} & \textbf{38.4} & \textbf{37.0} & 35.1 & \textbf{31.0} & \textbf{67.2} & \textbf{66.8}$^{\text{ns}}$ & 65.1 & \textbf{61.3} & \textbf{52.9} & 95.3 & 95.3$^{\text{ns}}$ & 95.0 & \textbf{94.3} & \textbf{92.4} & \textbf{62.5} & \textbf{62.3}$^{\text{ns}}$ & 60.4 & \textbf{57.4} & \textbf{51.3} \\
 \hline
\end{tabular}
\end{table*}

\subsubsection{Evaluation Protocol}
We evaluate all models at target dimensions $k \in \{768, 512, 256, 128, 64\}$. For models with a native dimension of 768,  the $k=768$ case coincides with no dimensionality reduction.
We report \textbf{MRR@10} for MS MARCO and \textbf{nDCG@10} for the remaining datasets, following standard conventions.

\subsubsection{Implementation Details}
All spectral decompositions and transformations are implemented in NumPy. Embeddings are generated using the original model checkpoints with default configurations. The covariance matrix and noise-floor statistics are estimated from the document corpus of each dataset, using up to 1M randomly sampled documents or the full corpus when fewer are available.  All experiments were conducted on Snellius, the Dutch National Supercomputer, using NVIDIA H100 GPUs.

\subsection{Experiment Results}

We organize the results into four parts. 
We first report the main retrieval results on three representative models, then evaluate the consistency of the observed trends on additional retrievers. 
Next, we assess how well the predicted tempering exponent aligns with the empirical optimum, and finally examine the sensitivity of the method to the choice of tail set~$\mathcal{T}$.

\subsubsection{Main Results}

We focus our main analysis on three representative models (Qwen3-8B, Jina-v4, GTE-7B) covering diverse architectures and scales. 
As shown in Table~\ref{tab:main_results}, our SpecTemp method achieves the best or tied-best performance among spectral methods in the majority of configurations without any tuning. 
PCA performs well at high dimensions but degrades under aggressive compression, while Whitening shows the opposite pattern; the fixed $\gamma$-Whitening offers a compromise but cannot adapt across compression regimes. SpecTemp automatically adjusts its tempering exponent and consistently matches or outperforms all fixed-$\gamma$ alternatives. 
Prefix Truncation is competitive on FEVER for MRL-trained models but falls behind on non-MRL models and on tasks with complex query semantics (e.g., FiQA), as it is restricted to the first $k$ training-time coordinates. Spectral methods—and SpecTemp in particular—consistently lead in these settings by projecting onto corpus-adaptive eigenvectors with richer expressivity.

\subsubsection{Consistency across Retrieval Models}

 \begin{figure}[t]
    \centering
    \includegraphics[width=\linewidth]{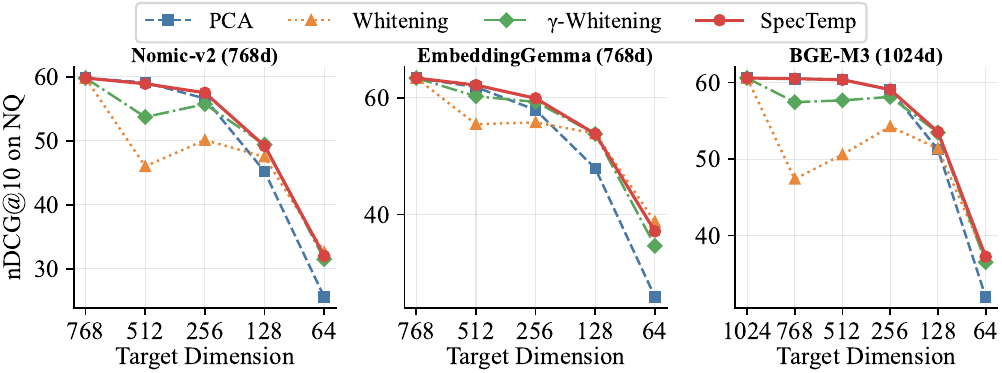}
\caption{Performance consistency on NQ across additional retrievers. BGE-M3 has 1024 native dimensions, while the others have 768.}
    \label{fig:generalization}
\end{figure}
To verify that our findings generalize beyond the main evaluation, we test on three additional models on the NQ dataset: Nomic-v2, EmbeddingGemma, and BGE-M3 (Figure~\ref{fig:generalization}).
We compare spectral methods only, as they share the same eigendecomposition backbone and isolate the effect of tempering strategy. The results reveal a consistent pattern: PCA degrades sharply at low dimensions where its skewed energy distribution fails to preserve fine distinctions, while Whitening suffers at high dimensions where it amplifies spectral noise. SpecTemp remains on the Pareto frontier across all dimensions, confirming that the adaptive $\gamma(k)$ mechanism robustly balances signal preservation and noise suppression across diverse architectures and scales.

\subsubsection{Alignment with Empirical Optima}

\begin{table}[t]
    \centering
 
    \caption{Comparison between oracle $\gamma^*(k)$ obtained via grid search and the theoretically predicted $\gamma(k)$ on NQ by GTE-7B.}
    \label{tab:gamma_alignment}
    \setlength{\tabcolsep}{5pt} 
    \renewcommand{\arraystretch}{1.1} %
    \begin{tabular}{lccccc}
        \toprule
        \textbf{Target Dimension} $k$ ($\rightarrow$) & \textbf{768} & \textbf{512} & \textbf{256} & \textbf{128} & \textbf{64} \\
        \midrule
        Oracle $\gamma^*_{\text{grid}}$ (Empirical)   &0.15   &0.25  &0.45   &0.55   &0.95  \\
        Predicted $\gamma(k)$ (SpecTemp) & 0.15 & 0.24  & 0.49  & 0.96   &1.00   \\
        $|\Delta|$ nDCG@10 \scriptsize{(0--100 scale)} &0.02   & 0.01 &0.06 &0.11 & 0.05  \\
        \bottomrule
    \end{tabular}
\end{table}

To validate that our predicted $\gamma(k)$ tracks the true optimum, we perform a grid search over $\gamma \in \{0, 0.05, \ldots, 1.0\}$ on GTE-7B (NQ), selecting the best-performing $\gamma$ at each target dimension. 
As shown in Table~\ref{tab:gamma_alignment}, the predicted $\gamma(k)$ closely matches the oracle at most dimensions. 
At $k{=}128$, despite the divergence in parameter space ($0.55$ vs.\ $0.96$), the resulting performance penalty is minimal ($|\Delta|\text{nDCG@10}=0.11$ points on a 0--100 scale).
This indicates a flat optimization landscape where SpecTemp successfully locates a robust operating point within the near-optimal basin, achieving near-oracle performance without expensive validation, with an average $|\Delta|$ of just 0.05 points.

\subsubsection{Sensitivity Analysis of~$\mathcal{T}$}\label{sec:sensitivity_analysis}

We test sensitivity to the tail set $\mathcal{T}$ in Eq.~\ref{eq:noise_floor} by varying its size from 5\% to 20\%.
On GTE-7B \(\rightarrow\) NQ, nDCG@10 varies by at most 0.03 on a 0–100 scale.
Given this robustness, we set $\mathcal{T}$ to the last 10\% of eigenvalue indices for all experiments without per-task tuning. This confirms that the noise floor estimate is stable across percentiles and requires no calibration. \looseness=-1

\section{Conclusion}
We proposed  \textbf{SpecTemp}, a learning-free post-hoc compression method for dense retrieval embeddings. 
By deriving a dimensionality-adaptive tempering exponent $\gamma(k)$ from the local SNR profile of the eigenspectrum, our method effectively bridges the trade-off between variance preservation (PCA) and isotropy (Whitening). 
Extensive experiments across six diverse models show that SpecTemp closely matches grid-searched oracle $\gamma^*(k)$ performance without any hyperparameter tuning, making it a practical baseline for learning-free embedding compression.

Beyond learning-free post-hoc linear compression, future work could examine how SpecTemp interacts with orthogonal index-side compression methods such as PQ, as well as its impact on end-to-end deployment metrics beyond retrieval effectiveness alone.

\begin{acks}
This work was partially supported by the China Scholarship Council (202308440220), the LESSEN project (NWA.1389.20.183) of the research program NWA ORC 2020/21 which is financed by the Dutch Research Council (NWO), and the PACINO project~(215742) which is financed by the Swiss National Science Foundation (SNSF). 
The authors acknowledge the peoples of the Woi Wurrung and Boon Wurrung language groups of the eastern Kulin Nation on whose unceded lands ACM SIGIR 2026 was hosted. We pay our respects to their Elders past and present, and extend that respect to all Aboriginal and Torres Strait Islander peoples today and their continuing connection to land, sea, sky, and community.
\end{acks}

\bibliographystyle{ACM-Reference-Format}
\balance
\bibliography{main}

@inproceedings{KarpukhinOMLWEC20_DPR,
  author       = {Vladimir Karpukhin and
                  Barlas Oguz and
                  Sewon Min and
                  Patrick S. H. Lewis and
                  Ledell Wu and
                  Sergey Edunov and
                  Danqi Chen and
                  Wen{-}tau Yih},
  title        = {Dense Passage Retrieval for Open-Domain Question Answering},
  booktitle    = {Proceedings of the 2020 Conference on Empirical Methods in Natural
                  Language Processing, {EMNLP} 2020},
  pages        = {6769--6781},
  publisher    = {Association for Computational Linguistics},
  year         = {2020},
  doi          = {10.18653/V1/2020.EMNLP-MAIN.550},
  bibsource    = {dblp computer science bibliography, https://dblp.org}
}

@inproceedings{XiongXLTLBAO21_ANCE,
  author       = {Lee Xiong and
                  Chenyan Xiong and
                  Ye Li and
                  Kwok{-}Fung Tang and
                  Jialin Liu and
                  Paul N. Bennett and
                  Junaid Ahmed and
                  Arnold Overwijk},
  title        = {Approximate Nearest Neighbor Negative Contrastive Learning for Dense
                  Text Retrieval},
  booktitle    = {9th International Conference on Learning Representations, {ICLR} 2021,
                  Virtual Event, Austria, May 3-7, 2021},
  publisher    = {OpenReview.net},
  year         = {2021},
  url          = {https://openreview.net/forum?id=zeFrfgyZln},
  bibsource    = {dblp computer science bibliography, https://dblp.org}
}

@inproceedings{ZhongHWC23_Poisoning,
  author       = {Zexuan Zhong and
                  Ziqing Huang and
                  Alexander Wettig and
                  Danqi Chen},
  title        = {Poisoning Retrieval Corpora by Injecting Adversarial Passages},
  booktitle    = {Proceedings of the 2023 Conference on Empirical Methods in Natural
                  Language Processing, {EMNLP} 2023, Singapore, December 6-10, 2023},
  pages        = {13764--13775},
  publisher    = {Association for Computational Linguistics},
  year         = {2023},
  url          = {https://doi.org/10.18653/v1/2023.emnlp-main.849},
  doi          = {10.18653/V1/2023.EMNLP-MAIN.849},
  bibsource    = {dblp computer science bibliography, https://dblp.org}
}

@article{bajaj2016ms,
title={Ms marco: A human generated machine reading comprehension dataset},
author={Bajaj, Payal and Campos, Daniel and Craswell, Nick and Deng, Li and Gao, Jianfeng and Liu, Xiaodong and Majumder, Rangan and McNamara, Andrew and Mitra, Bhaskar and Nguyen, Tri and others},
journal={arXiv preprint arXiv:1611.09268},
year={2016}
}

@article{KwiatkowskiPRCP19_NQ,
  author       = {Tom Kwiatkowski and
                  Jennimaria Palomaki and
                  Olivia Redfield and
                  Michael Collins and
                  Ankur P. Parikh and
                  Chris Alberti and
                  Danielle Epstein and
                  Illia Polosukhin and
                  Jacob Devlin and
                  Kenton Lee and
                  Kristina Toutanova and
                  Llion Jones and
                  Matthew Kelcey and
                  Ming{-}Wei Chang and
                  Andrew M. Dai and
                  Jakob Uszkoreit and
                  Quoc Le and
                  Slav Petrov},
  title        = {Natural Questions: a Benchmark for Question Answering Research},
  journal      = {Trans. Assoc. Comput. Linguistics},
  volume       = {7},
  pages        = {452--466},
  year         = {2019},
  url          = {https://doi.org/10.1162/tacl\_a\_00276},
  doi          = {10.1162/TACL\_A\_00276},
  bibsource    = {dblp computer science bibliography, https://dblp.org}
}

@inproceedings{HofstatterLYLH21_tasb_dense_retrieval,
  author       = {Sebastian Hofst{\"{a}}tter and
                  Sheng{-}Chieh Lin and
                  Jheng{-}Hong Yang and
                  Jimmy Lin and
                  Allan Hanbury},
  title        = {Efficiently Teaching an Effective Dense Retriever with Balanced Topic Aware Sampling},
  booktitle    = {{SIGIR} '21: The 44th International {ACM} {SIGIR} Conference on Research  and Development in Information Retrieval, Virtual Event, Canada, July 11-15, 2021},
  pages        = {113--122},
  publisher    = {{ACM}},
  year         = {2021},
  url          = {https://doi.org/10.1145/3404835.3462891},
  doi          = {10.1145/3404835.3462891},
  bibsource    = {dblp computer science bibliography, https://dblp.org}
}

@inproceedings{LinALOLMY023_dragon,
  author       = {Sheng{-}Chieh Lin and
                  Akari Asai and
                  Minghan Li and
                  Barlas Oguz and
                  Jimmy Lin and
                  Yashar Mehdad and
                  Wen{-}tau Yih and
                  Xilun Chen},
  title        = {How to Train Your Dragon: Diverse Augmentation Towards Generalizable Dense Retrieval},
  booktitle    = {Findings of the Association for Computational Linguistics: {EMNLP}   2023, Singapore, December 6-10, 2023},
  pages        = {6385--6400},
  publisher    = {Association for Computational Linguistics},
  year         = {2023},
  url          = {https://doi.org/10.18653/v1/2023.findings-emnlp.423},
  doi          = {10.18653/V1/2023.FINDINGS-EMNLP.423},
  bibsource    = {dblp computer science bibliography, https://dblp.org}
}

@inproceedings{10.1145/3583780.3614793_MCARA,
author = {Liu, Yu-An and Zhang, Ruqing and Guo, Jiafeng and de Rijke, Maarten and Chen, Wei and Fan, Yixing and Cheng, Xueqi},
title = {Black-box Adversarial Attacks against Dense Retrieval Models: A Multi-view Contrastive Learning Method},
year = {2023},
isbn = {9798400701245},
publisher = {Association for Computing Machinery},
address = {New York, NY, USA},
url = {https://doi.org/10.1145/3583780.3614793},
doi = {10.1145/3583780.3614793},
booktitle = {Proceedings of the 32nd ACM International Conference on Information and Knowledge Management},
pages = {1647–1656},
numpages = {10},
location = {<conf-loc>, <city>Birmingham</city>, <country>United Kingdom</country>, </conf-loc>},
series = {CIKM '23}
}

@inproceedings{li2025reproducinghotflip,
  author       = {Yongkang Li and
                  Panagiotis Eustratiadis and
                  Evangelos Kanoulas},
  title        = {Reproducing HotFlip for Corpus Poisoning Attacks in Dense Retrieval},
  booktitle    = {Advances in Information Retrieval - 47th European Conference on Information
                  Retrieval, {ECIR} 2025, Lucca, Italy, April 6-10, 2025, Proceedings,
                  Part {IV}},
  series       = {Lecture Notes in Computer Science},
  volume       = {15575},
  pages        = {95--111},
  publisher    = {Springer},
  year         = {2025},
  url          = {https://doi.org/10.1007/978-3-031-88717-8\_8},
  doi          = {10.1007/978-3-031-88717-8\_8},
  bibsource    = {dblp computer science bibliography, https://dblp.org}
}

@article{WangE5-base,
  author       = {Liang Wang and
                  Nan Yang and
                  Xiaolong Huang and
                  Binxing Jiao and
                  Linjun Yang and
                  Daxin Jiang and
                  Rangan Majumder and
                  Furu Wei},
  title        = {Text Embeddings by Weakly-Supervised Contrastive Pre-training},
  journal      = {CoRR},
  volume       = {abs/2212.03533},
  year         = {2022},
  url          = {https://doi.org/10.48550/arXiv.2212.03533},
  doi          = {10.48550/ARXIV.2212.03533},
  eprinttype    = {arXiv},
  eprint       = {2212.03533},
  bibsource    = {dblp computer science bibliography, https://dblp.org}
}

@inproceedings{MaWYWL24_RankLLaMA,
  author       = {Xueguang Ma and
                  Liang Wang and
                  Nan Yang and
                  Furu Wei and
                  Jimmy Lin},
 
  title        = {Fine-Tuning LLaMA for Multi-Stage Text Retrieval},
  booktitle    = {Proceedings of the 47th International {ACM} {SIGIR} Conference on
                  Research and Development in Information Retrieval, {SIGIR} 2024, Washington
                  DC, USA, July 14-18, 2024},
  pages        = {2421--2425},
  publisher    = {{ACM}},
  year         = {2024},
  url          = {https://doi.org/10.1145/3626772.3657951},
  doi          = {10.1145/3626772.3657951},
  bibsource    = {dblp computer science bibliography, https://dblp.org}
}

@inproceedings{MaiaHFDMZB18_fiqa,
  author       = {Macedo Maia and
                  Siegfried Handschuh and
                  Andr{\'{e}} Freitas and
                  Brian Davis and
                  Ross McDermott and
                  Manel Zarrouk and
                  Alexandra Balahur},
 
  title        = {WWW'18 Open Challenge: Financial Opinion Mining and Question Answering},
  booktitle    = {Companion of the The Web Conference 2018 on The Web Conference 2018,
                  {WWW} 2018, Lyon , France, April 23-27, 2018},
  pages        = {1941--1942},
  publisher    = {{ACM}},
  year         = {2018},
  url          = {https://doi.org/10.1145/3184558.3192301},
  bibsource    = {dblp computer science bibliography, https://dblp.org}
}

@misc{zhang2025qwen3embedding,
      title={Qwen3 Embedding: Advancing Text Embedding and Reranking Through Foundation Models}, 
      author={Yanzhao Zhang and Mingxin Li and Dingkun Long and Xin Zhang and Huan Lin and Baosong Yang and Pengjun Xie and An Yang and Dayiheng Liu and Junyang Lin and Fei Huang and Jingren Zhou},
      year={2025},
      eprint={2506.05176},
      archivePrefix={arXiv},
      primaryClass={cs.CL},
      url={https://arxiv.org/abs/2506.05176}, 
}

@misc{nussbaum2025_nomicv2,
      title={Training Sparse Mixture Of Experts Text Embedding Models}, 
      author={Zach Nussbaum and Brandon Duderstadt},
      year={2025},
      eprint={2502.07972},
      archivePrefix={arXiv},
      primaryClass={cs.CL},
}

@misc{vera2025embeddinggemma,
      title={EmbeddingGemma: Powerful and Lightweight Text Representations}, 
      author={Henrique Schechter Vera and Sahil Dua and Biao Zhang and Daniel Salz and Ryan Mullins and Sindhu Raghuram Panyam and Sara Smoot and Iftekhar Naim and Joe Zou and Feiyang Chen and Daniel Cer and Alice Lisak and Min Choi and Lucas Gonzalez and Omar Sanseviero and Glenn Cameron and Ian Ballantyne and Kat Black and Kaifeng Chen and Weiyi Wang and Zhe Li and Gus Martins and Jinhyuk Lee and Mark Sherwood and Juyeong Ji and Renjie Wu and Jingxiao Zheng and Jyotinder Singh and Abheesht Sharma and Divyashree Sreepathihalli and Aashi Jain and Adham Elarabawy and AJ Co and Andreas Doumanoglou and Babak Samari and Ben Hora and Brian Potetz and Dahun Kim and Enrique Alfonseca and Fedor Moiseev and Feng Han and Frank Palma Gomez and Gustavo Hernández Ábrego and Hesen Zhang and Hui Hui and Jay Han and Karan Gill and Ke Chen and Koert Chen and Madhuri Shanbhogue and Michael Boratko and Paul Suganthan and Sai Meher Karthik Duddu and Sandeep Mariserla and Setareh Ariafar and Shanfeng Zhang and Shijie Zhang and Simon Baumgartner and Sonam Goenka and Steve Qiu and Tanmaya Dabral and Trevor Walker and Vikram Rao and Waleed Khawaja and Wenlei Zhou and Xiaoqi Ren and Ye Xia and Yichang Chen and Yi-Ting Chen and Zhe Dong and Zhongli Ding and Francesco Visin and Gaël Liu and Jiageng Zhang and Kathleen Kenealy and Michelle Casbon and Ravin Kumar and Thomas Mesnard and Zach Gleicher and Cormac Brick and Olivier Lacombe and Adam Roberts and Qin Yin and Yunhsuan Sung and Raphael Hoffmann and Tris Warkentin and Armand Joulin and Tom Duerig and Mojtaba Seyedhosseini},
      year={2025},
      eprint={2509.20354},
      archivePrefix={arXiv},
      primaryClass={cs.CL},
}

@misc{long2025diver,
      title={DIVER: A Multi-Stage Approach for Reasoning-intensive Information Retrieval}, 
      author={Meixiu Long and Duolin Sun and Dan Yang and Junjie Wang and Yue Shen and Jian Wang and Peng Wei and Jinjie Gu and Jiahai Wang},
      year={2025},
      eprint={2508.07995},
      archivePrefix={arXiv},
      primaryClass={cs.IR},
}

@inproceedings{KusupatiBRWSRHC22_MRL,
  author       = {Aditya Kusupati and
                  Gantavya Bhatt and
                  Aniket Rege and
                  Matthew Wallingford and
                  Aditya Sinha and
                  Vivek Ramanujan and
                  William Howard{-}Snyder and
                  Kaifeng Chen and
                  Sham M. Kakade and
                  Prateek Jain and
                  Ali Farhadi},
  title        = {Matryoshka Representation Learning},
  booktitle    = {Advances in Neural Information Processing Systems 35: Annual Conference
                  on Neural Information Processing Systems 2022, NeurIPS 2022, New Orleans,
                  LA, USA, November 28 - December 9, 2022},
  year         = {2022},
  url          = {http://papers.nips.cc/paper\_files/paper/2022/hash/c32319f4868da7613d78af9993100e42-Abstract-Conference.html},
  bibsource    = {dblp computer science bibliography, https://dblp.org}
}

@misc{günther2025jinaembeddingsv4,
      title={jina-embeddings-v4: Universal Embeddings for Multimodal Multilingual Retrieval}, 
      author={Michael Günther and Saba Sturua and Mohammad Kalim Akram and Isabelle Mohr and Andrei Ungureanu and Sedigheh Eslami and Scott Martens and Bo Wang and Nan Wang and Han Xiao},
      year={2025},
      eprint={2506.18902},
      archivePrefix={arXiv},
      primaryClass={cs.AI},
      url={https://arxiv.org/abs/2506.18902}, 
}

@misc{li2023gte_7b,
      title={Towards General Text Embeddings with Multi-stage Contrastive Learning}, 
      author={Zehan Li and Xin Zhang and Yanzhao Zhang and Dingkun Long and Pengjun Xie and Meishan Zhang},
      year={2023},
      eprint={2308.03281},
      archivePrefix={arXiv},
      primaryClass={cs.CL},
      url={https://arxiv.org/abs/2308.03281}, 
}

@inproceedings{takeshita_2025_randomly50,
    title = "Randomly Removing 50{\%} of Dimensions in Text Embeddings has Minimal Impact on Retrieval and Classification Tasks",
    author = "Takeshita, Sotaro  and
      Takeshita, Yurina  and
      Ruffinelli, Daniel  and
      Ponzetto, Simone Paolo",
    booktitle = "Proceedings of the 2025 Conference on Empirical Methods in Natural Language Processing",
    month = nov,
    year = "2025",
    address = "Suzhou, China",
    publisher = "Association for Computational Linguistics",
    url = "https://aclanthology.org/2025.emnlp-main.1410/",
    doi = "10.18653/v1/2025.emnlp-main.1410",
    pages = "27705--27726",
    ISBN = "979-8-89176-332-6"
}

@inproceedings{ SatopaaAIR11_Kneedle,
  author       = {Ville Satopaa and
                  Jeannie R. Albrecht and
                  David E. Irwin and
                  Barath Raghavan},
  title        = {Finding a "Kneedle" in a Haystack: Detecting Knee Points in System
                  Behavior},
  booktitle    = {31st {IEEE} International Conference on Distributed Computing Systems
                  Workshops {(ICDCS} 2011 Workshops), 20-24 June 2011, Minneapolis,
                  Minnesota, {USA}},
  pages        = {166--171},
  publisher    = {{IEEE} Computer Society},
  year         = {2011},
  url          = {https://doi.org/10.1109/ICDCSW.2011.20},
  doi          = {10.1109/ICDCSW.2011.20},
  bibsource    = {dblp computer science bibliography, https://dblp.org}
}

@inproceedings{ma-etal-2021-simple,
    title = "Simple and Effective Unsupervised Redundancy Elimination to Compress Dense Vectors for Passage Retrieval",
    author = "Ma, Xueguang  and
      Li, Minghan  and
      Sun, Kai  and
      Xin, Ji  and
      Lin, Jimmy",
    booktitle = "Proceedings of the 2021 Conference on Empirical Methods in Natural Language Processing",
    month = nov,
    year = "2021",
    address = "Online and Punta Cana, Dominican Republic",
    publisher = "Association for Computational Linguistics",
    url = "https://aclanthology.org/2021.emnlp-main.227/",
    pages = "2854--2859",
}

@misc{su2021whiteningsentence,
      title={Whitening Sentence Representations for Better Semantics and Faster Retrieval}, 
      author={Jianlin Su and Jiarun Cao and Weijie Liu and Yangyiwen Ou},
      year={2021},
      eprint={2103.15316},
      archivePrefix={arXiv},
      primaryClass={cs.CL},
      url={https://arxiv.org/abs/2103.15316}, 
}

@inproceedings{huang-etal-2021-whiteningbert-easy,
    title = "{W}hitening{BERT}: An Easy Unsupervised Sentence Embedding Approach",
    author = "Huang, Junjie  and
      Tang, Duyu  and
      Zhong, Wanjun  and
      Lu, Shuai  and
      Shou, Linjun  and
      Gong, Ming  and
      Jiang, Daxin  and
      Duan, Nan",
    booktitle = "Findings of the Association for Computational Linguistics: EMNLP 2021",
    month = nov,
    year = "2021",
    address = "Punta Cana, Dominican Republic",
    publisher = "Association for Computational Linguistics",
    url = "https://aclanthology.org/2021.findings-emnlp.23/",
    pages = "238--244"
}

@article{johnson1984extensions,
  title={Extensions of Lipschitz mappings into a Hilbert space},
  author={Johnson, William B and Lindenstrauss, Joram and others},
  journal={Contemporary mathematics},
  volume={26},
  number={189-206},
  pages={1},
  year={1984},
    url={https://api.semanticscholar.org/CorpusID:117819162}
}

@inproceedings{
mu2018allbutthetop,
title={All-but-the-Top: Simple and Effective Postprocessing for Word Representations},
author={Jiaqi Mu and Pramod Viswanath},
booktitle={International Conference on Learning Representations},
year={2018},
url={https://openreview.net/forum?id=HkuGJ3kCb},
}

@inproceedings{rajaee-pilehvar-2021-cluster,
    title = "A Cluster-based Approach for Improving Isotropy in Contextual Embedding Space",
    author = "Rajaee, Sara  and
      Pilehvar, Mohammad Taher",
    booktitle = "Proceedings of the 59th Annual Meeting of the Association for Computational Linguistics and the 11th International Joint Conference on Natural Language Processing (Volume 2: Short Papers)",
    month = aug,
    year = "2021",
    address = "Online",
    publisher = "Association for Computational Linguistics",
    url = "https://aclanthology.org/2021.acl-short.73/",
    pages = "575--584",
}

@inproceedings{raunak-etal-2019-effective,
    title = "Effective Dimensionality Reduction for Word Embeddings",
    author = "Raunak, Vikas  and
      Gupta, Vivek  and
      Metze, Florian",
    booktitle = "Proceedings of the 4th Workshop on Representation Learning for NLP (RepL4NLP-2019)",
    month = aug,
    year = "2019",
    address = "Florence, Italy",
    publisher = "Association for Computational Linguistics",
    url = "https://aclanthology.org/W19-4328/",
    pages = "235--243",
}

@article{jegou2010product,
  author       = {Herv{\'{e}} J{\'{e}}gou and
                  Matthijs Douze and
                  Cordelia Schmid},
  title        = {Product Quantization for Nearest Neighbor Search},
  journal      = {{IEEE} Trans. Pattern Anal. Mach. Intell.},
  volume       = {33},
  number       = {1},
  pages        = {117--128},
  year         = {2011},
  url          = {https://doi.org/10.1109/TPAMI.2010.57},
  doi          = {10.1109/TPAMI.2010.57},
  bibsource    = {dblp computer science bibliography, https://dblp.org}
}

@inproceedings{ DevlinCLT19_bert,
  author       = {Jacob Devlin and
                  Ming{-}Wei Chang and
                  Kenton Lee and
                  Kristina Toutanova},
  title        = {{BERT:} Pre-training of Deep Bidirectional Transformers for Language
                  Understanding},
  booktitle    = {Proceedings of the 2019 Conference of the North American Chapter of
                  the Association for Computational Linguistics: Human Language Technologies,
                  {NAACL-HLT} 2019, Minneapolis, MN, USA, June 2-7, 2019, Volume 1 (Long
                  and Short Papers)},
  pages        = {4171--4186},
  publisher    = {Association for Computational Linguistics},
  year         = {2019},
  doi          = {10.18653/V1/N19-1423},
  bibsource    = {dblp computer science bibliography, https://dblp.org}
}

@online{kexuefm-9079,
  title   = {When BERT Whitening Introduces Hyperparameters: There Is Always One That Suits You},
  author  = {Su, Jianlin},
  year    = {2022},
  month   = {May},
  url     = {https://kexue.fm/archives/9079},
  note    = {Chinese blog post}
}

@misc{zuo2026efficiencyembeddingcompress ,
      title={More Than Efficiency: Embedding Compression Improves Domain Adaptation in Dense Retrieval}, 
      author={Chunsheng Zuo and Daniel Khashabi},
      year={2026},
      eprint={2601.13525},
      archivePrefix={arXiv},
      primaryClass={cs.IR},
      url={https://arxiv.org/abs/2601.13525}, 
}

@inproceedings{li-etal-2020-sentence,
    title = "On the Sentence Embeddings from Pre-trained Language Models",
    author = "Li, Bohan  and
      Zhou, Hao  and
      He, Junxian  and
      Wang, Mingxuan  and
      Yang, Yiming  and
      Li, Lei",
    booktitle = "Proceedings of the 2020 Conference on Empirical Methods in Natural Language Processing (EMNLP)",
    month = nov,
    year = "2020",
    address = "Online",
    publisher = "Association for Computational Linguistics",
    url = "https://aclanthology.org/2020.emnlp-main.733/",
    pages = "9119--9130",
}

@inproceedings{ ThorneVCM18_FEVER,
  author       = {James Thorne and
                  Andreas Vlachos and
                  Christos Christodoulopoulos and
                  Arpit Mittal},
  title        = {{FEVER:} a Large-scale Dataset for Fact Extraction and VERification},
  booktitle    = {Proceedings of the 2018 Conference of the North American Chapter of
                  the Association for Computational Linguistics: Human Language Technologies,
                  {NAACL-HLT} 2018, New Orleans, Louisiana, USA, June 1-6, 2018, Volume
                  1 (Long Papers)},
  pages        = {809--819},
  publisher    = {Association for Computational Linguistics},
  year         = {2018},
  url          = {https://doi.org/10.18653/v1/n18-1074},
  doi          = {10.18653/V1/N18-1074},
  bibsource    = {dblp computer science bibliography, https://dblp.org}
}

@inproceedings{chen-etal-2024-m3,
    title = "{M}3-Embedding: Multi-Linguality, Multi-Functionality, Multi-Granularity Text Embeddings Through Self-Knowledge Distillation",
    author = "Chen, Jianlyu  and
      Xiao, Shitao  and
      Zhang, Peitian  and
      Luo, Kun  and
      Lian, Defu  and
      Liu, Zheng",
    booktitle = "Findings of the Association for Computational Linguistics: ACL 2024",
    month = aug,
    year = "2024",
    address = "Bangkok, Thailand",
    publisher = "Association for Computational Linguistics",
    url = "https://aclanthology.org/2024.findings-acl.137/",
    doi = "10.18653/v1/2024.findings-acl.137",
    pages = "2318--2335"
}

@inproceedings{reimers-2019-sentence-bert,
  author       = {Nils Reimers and
                  Iryna Gurevych},
  title        = {Sentence-BERT: Sentence Embeddings using Siamese BERT-Networks},
  booktitle    = {Proceedings of the 2019 Conference on Empirical Methods in Natural
                  Language Processing and the 9th International Joint Conference on
                  Natural Language Processing, {EMNLP-IJCNLP} 2019, Hong Kong, China,
                  November 3-7, 2019},
  pages        = {3980--3990},
  publisher    = {Association for Computational Linguistics},
  year         = {2019},
  url          = {https://doi.org/10.18653/v1/D19-1410},
  doi          = {10.18653/V1/D19-1410},
  bibsource    = {dblp computer science bibliography, https://dblp.org}
}

@inproceedings{liu-etal-2022-dimension,
    title = "Dimension Reduction for Efficient Dense Retrieval via Conditional Autoencoder",
    author = "Liu, Zhenghao  and
      Zhang, Han  and
      Xiong, Chenyan  and
      Liu, Zhiyuan  and
      Gu, Yu  and
      Li, Xiaohua",
    booktitle = "Proceedings of the 2022 Conference on Empirical Methods in Natural Language Processing",
    month = dec,
    year = "2022",
    address = "Abu Dhabi, United Arab Emirates",
    publisher = "Association for Computational Linguistics",
    url = "https://aclanthology.org/2022.emnlp-main.384/",
    doi = "10.18653/v1/2022.emnlp-main.384",
    pages = "5692--5698"
}

@article{douze2024faiss,
      title={The Faiss library},
      author={Matthijs Douze and Alexandr Guzhva and Chengqi Deng and Jeff Johnson and Gergely Szilvasy and Pierre-Emmanuel Mazaré and Maria Lomeli and Lucas Hosseini and Hervé Jégou},
      year={2024},
      eprint={2401.08281},
      archivePrefix={arXiv},
      primaryClass={cs.LG}
}

@inproceedings{lioutas-etal-2020-improving,
    title = "{I}mproving {W}ord {E}mbedding {F}actorization for {C}ompression {U}sing {D}istilled {N}onlinear {N}eural {D}ecomposition",
    author = "Lioutas, Vasileios  and
      Rashid, Ahmad  and
      Kumar, Krtin  and
      Haidar, Md. Akmal  and
      Rezagholizadeh, Mehdi",
    booktitle = "Findings of the Association for Computational Linguistics: EMNLP 2020",
    month = nov,
    year = "2020",
    address = "Online",
    publisher = "Association for Computational Linguistics",
    url = "https://aclanthology.org/2020.findings-emnlp.250/",
    doi = "10.18653/v1/2020.findings-emnlp.250",
    pages = "2774--2784"
}

@misc{zhang2026caseconditionawaresentence,
      title={CASE -- Condition-Aware Sentence Embeddings for Conditional Semantic Textual Similarity Measurement}, 
      author={Gaifan Zhang and Yi Zhou and Danushka Bollegala},
      year={2026},
      eprint={2503.17279},
      archivePrefix={arXiv},
      primaryClass={cs.CL},
      url={https://arxiv.org/abs/2503.17279}, 
}

@inproceedings{zhang-etal-2024-evaluating-unsupervised,
    title = "Evaluating Unsupervised Dimensionality Reduction Methods for Pretrained Sentence Embeddings",
    author = "Zhang, Gaifan  and
      Zhou, Yi  and
      Bollegala, Danushka",
    booktitle = "Proceedings of the 2024 Joint International Conference on Computational Linguistics, Language Resources and Evaluation (LREC-COLING 2024)",
    month = may,
    year = "2024",
    address = "Torino, Italia",
    publisher = "ELRA and ICCL",
    url = "https://aclanthology.org/2024.lrec-main.579/",
    pages = "6530--6543"
}

@inproceedings{li2026understanding,
  author       = {Yongkang Li},
  title        = {Understanding and Enhancing Robustness in Dense Information Retrieval},
  booktitle    = {Advances in Information Retrieval - 48th European Conference on Information
                  Retrieval, {ECIR} 2026, Delft, The Netherlands, March 29 - April 2,
                  2026, Proceedings, Part {III}},
  series       = {Lecture Notes in Computer Science},
  pages        = {599--607},
  publisher    = {Springer},
  year         = {2026},
  url          = {https://doi.org/10.1007/978-3-032-21324-2\_51},
  doi          = {10.1007/978-3-032-21324-2\_51},
  bibsource    = {dblp computer science bibliography, https://dblp.org}
}

@inproceedings{li2025unsupervised,
  author       = {Yongkang Li and
                  Panagiotis Eustratiadis and
                  Simon Lupart and
                  Evangelos Kanoulas},
  title        = {Unsupervised Corpus Poisoning Attacks in Continuous Space for Dense
                  Retrieval},
  booktitle    = {Proceedings of the 48th International {ACM} {SIGIR} Conference on
                  Research and Development in Information Retrieval, {SIGIR} 2025, Padua,
                  Italy, July 13-18, 2025},
  pages        = {2452--2462},
  publisher    = {{ACM}},
  year         = {2025},
  url          = {https://doi.org/10.1145/3726302.3730110},
  doi          = {10.1145/3726302.3730110},
  bibsource    = {dblp computer science bibliography, https://dblp.org}
}

@inproceedings{ PenhaCH22_query_variation,
  author       = {Gustavo Penha and
                  Arthur C{\^{a}}mara and
                  Claudia Hauff},
 
  title        = {Evaluating the Robustness of Retrieval Pipelines with Query Variation Generators},
  booktitle    = {Advances in Information Retrieval - 44th European Conference on {IR}
                  Research, {ECIR} 2022, Stavanger, Norway, April 10-14, 2022, Proceedings,
                  Part {I}},
  series       = {Lecture Notes in Computer Science},
  volume       = {13185},
  pages        = {397--412},
  publisher    = {Springer},
  year         = {2022},
 
  doi          = {10.1007/978-3-030-99736-6\_27},
  bibsource    = {dblp computer science bibliography, https://dblp.org}
}


\end{document}